\documentclass[journal]{IEEEtran}
\usepackage{cite,url,subfigure,epsfig,graphicx}
\usepackage{tipa}
\usepackage{makecell}
\usepackage{multirow}
\usepackage{bbm}
\usepackage{mathrsfs}
\usepackage{array}
\usepackage{amssymb,amsmath}
\usepackage{indentfirst}
\usepackage{pifont}
\usepackage{algorithmic}
\usepackage{algorithm}
\usepackage{cases}
\usepackage{soul}
\usepackage{booktabs}
\usepackage[table]{xcolor}
\usepackage{graphicx}
\IEEEoverridecommandlockouts
\makeatletter
\usepackage{times,verbatim,amsfonts,color}
\usepackage{hyperref}
\makeatletter
\def\UrlAlphabet{%
      \do\a\do\b\do\c\do\d\do\e\do\f\do\g\do\h\do\i\do\j%
      \do\k\do\l\do\m\do\n\do\o\do\p\do\q\do\r\do\s\do\t%
      \do\u\do\v\do\w\do\x\do\y\do\z\do\A\do\B\do\C\do\D%
      \do\E\do\F\do\G\do\H\do\I\do\J\do\K\do\L\do\M\do\N%
      \do\O\do\P\do\Q\do\R\do\S\do\T\do\U\do\V\do\W\do\X%
      \do\Y\do\Z}
\def\UrlDigits{\do\1\do\2\do\3\do\4\do\5\do\6\do\7\do\8\do\9\do\0}
\g@addto@macro{\UrlBreaks}{\UrlOrds}
\g@addto@macro{\UrlBreaks}{\UrlAlphabet}
\g@addto@macro{\UrlBreaks}{\UrlDigits}
\makeatother

\usepackage{geometry}
\begin{document}

\title{Large Model Empowered Metaverse: State-of-the-Art, Challenges and Opportunities}

\author{
\IEEEauthorblockN{Yuntao~Wang\IEEEauthorrefmark{2}, Qinnan~Hu\IEEEauthorrefmark{2}, Zhou~Su\IEEEauthorrefmark{2}\IEEEauthorrefmark{1}, Linkang~Du\IEEEauthorrefmark{2}, Qichao~Xu\IEEEauthorrefmark{3}, and Weiwei Li\IEEEauthorrefmark{4}}\\
\IEEEauthorblockA{
\IEEEauthorrefmark{2}School of Cyber Science and Engineering, Xi'an Jiaotong University, China \\
\IEEEauthorrefmark{3}School of Mechatronic Engineering and Automation, Shanghai University, China\\
\IEEEauthorrefmark{4}College of Computer Science and Technology, Shanghai University of Electric Power, China\\
\IEEEauthorrefmark{1}Corresponding author: zhousu@ieee.org 
}}

\maketitle

\begin{abstract}
The Metaverse represents a transformative shift beyond traditional mobile Internet, creating an immersive, persistent digital ecosystem where users can interact, socialize, and work within 3D virtual environments. Powered by large models such as ChatGPT and Sora, the Metaverse benefits from precise large-scale real-world modeling, automated multimodal content generation, realistic avatars, and seamless natural language understanding, which enhance user engagement and enable more personalized, intuitive interactions. However, challenges remain, including limited scalability, constrained responsiveness, and low adaptability in dynamic environments. This paper investigates the integration of large models within the Metaverse, examining their roles in enhancing user interaction, perception, content creation, and service quality. To address existing challenges, we propose a generative AI-based framework for optimizing Metaverse rendering. This framework includes a cloud-edge-end collaborative model to allocate rendering tasks with minimal latency, a mobility-aware pre-rendering mechanism that dynamically adjusts to user movement, and a diffusion model-based adaptive rendering strategy to fine-tune visual details. Experimental results demonstrate the effectiveness of our approach in enhancing rendering efficiency and reducing rendering overheads, advancing large model deployment for a more responsive and immersive Metaverse.
\end{abstract}

\IEEEpeerreviewmaketitle

\section{Introduction}
\IEEEPARstart{T}{he} Metaverse, an immersive virtual space merging physical and digital realities, was first introduced in the 1990s by Neal Stephenson in his science fiction novel \textit{Snow Crash}. Recently, advances in virtual reality (VR), augmented reality (AR), artificial intelligence (AI), and blockchain have propelled the Metaverse's evolution from fiction to reality \cite{Wang2023MetaverseSurvey}. Recognized as the next-generation Internet paradigm, the Metaverse envisions a decentralized, immersive, and persistent digital ecosystem where users interact, socialize, create, shop, work, and play through their avatars. Beyond gaming and entertainment, the Metaverse spans diverse applications, including education, healthcare, commerce, and arts, offering new forms of digital living experiences \cite{Chamola2024Beyond}. The Metaverse also aims to establish an open, interconnected digital economy, enabling seamless transfer of assets and identities across various virtual platforms, fostering a diverse, inclusive, and innovation-driven digital society. This transformative digital landscape promises to reshape how people live, interact, and conduct business, ushering in a new era of interconnected, immersive experiences in a blended reality. As reported by P\&S Intelligence\footnotemark[1], the market size was valued at approximately \$129 billion in 2024 and is projected to grow at a compound annual growth rate (CAGR) of 44\%, potentially reaching \$1157 billion by 2030.
\footnotetext[1]{\url{https://www.psmarketresearch.com/market-analysis/metaverse-market}}

Large models, particularly large language models (LLMs) and large vision models (LVMs), characterized by their billions of parameters, have witnessed remarkable progress due to recent advancements in generative AI (GAI) algorithms, increased computational capabilities, and availability of Internet-scale datasets \cite{Wang2023ChatGPT}. Pioneering large models such as OpenAI's GPT-4o, Google's PaLM 2, Microsoft's Copilot, and OpenAI's Sora have established a foundation for creating highly sophisticated AI systems capable of understanding and generating human-like text, images, and other media.
In the Metaverse, large models play a transformative role, significantly enhancing its content richness, interaction immdersiveness, quality-of-experience (QoE), and quality-of-service (QoS) \cite{Chamola2024Beyond,Xu2023Generative,Saddik2024Integration}.
Tech giants including Microsoft, Google, OpenAI, Tencent, Meta, NVIDIA, and Alibaba are utilizing large models and GAI to enhance the Metaverse's capabilities, including virtual environments, digital avatars, interactive VR, 3D digital humans, digital twins, and immersive experiences.

Large models' powerful data processing and generation abilities contribute to generation of rich, multimodal content, including text, video, images, audio, and 3D interactive objects, which creates immersive and engaging virtual worlds \cite{Chamola2024Beyond}. They enable personalized avatars, interactive NPCs, and digital humans with lifelike behaviors and nuanced emotional expressions, enriching user interactions in virtual environments. 
Additionally, large models facilitate the construction of intricate virtual worlds by producing accurate digital twins, customized scenes, and context-aware feedback tailored to user interests \cite{Xu2023Generative,Saddik2024Integration}, which enhances engagement and QoE.
With their zero/few-shot learning and continuous learning capabilities \cite{Deepmind2024Scaling}, large models optimize Metaverse performance through intelligent resource allocation and adaptive rendering techniques, improving network efficiency and QoS \cite{Huynh2023AIMetaSurvey}. 
Despite these advancements, Metaverse systems powered by large models still face challenges: 1) limited scalability in virtual world construction {(i.e., how to support large-scale, immersive multi-user interactions under high computational cost of large model-driven scene rendering)}; 2) constrained responsiveness in rendering {(i.e., how to improve rendering responsiveness by integrating user mobility and intention prediction into real-time scene adaptation)}; 3) poor adaptability in dynamic environments {(i.e., how to enable personalized, fine-grained scene rendering that adapts to diverse user preferences, device capabilities, and dynamic interaction contexts)}. 

This paper provides a comprehensive review of recent advancements in large models empowered Metaverse system design and explores their integration across four key areas: interaction enhancement, multimodal perception, content generation, and QoE/QoS optimization. Key challenges faced by large model enabled Metaverse systems are identified, including scalability, adaptability, and responsiveness in dynamic virtual environments. To address these issues, we propose a GAI-based optimization approach for Metaverse rendering, comprising: 1) a cloud-edge-end collaborative model that efficiently allocates rendering tasks to reduce latency, 2) a mobility-aware pre-rendering mechanism that dynamically adjusts based on user movement and contextual cues, and 3) a diffusion model-based adaptive rendering strategy for customizing scene details based on user preferences and device capabilities. Experimental results demonstrate the feasibility and effectiveness of our approach, and we outline several crucial directions for future research in advancing large model applications in the Metaverse.

The remainder of this paper is organized as below. Section~\ref{SOTA} introduces the background and potential applications of large model empowered Metaverse and reviews state-of-the-art approaches.
Section~\ref{challenges} identifies key challenges in integrating large models and Metaverse. 
Section~\ref{solutions} presents a case study of GAI-based Metaverse rendering optimization to address these challenges.
Section~\ref{future} outlines future research directions. Finally, Section~\ref{conclusion} concludes this work.

\begin{figure}[!t]
\centering \setlength{\abovecaptionskip}{-0.cm}
  \includegraphics[width=9cm]{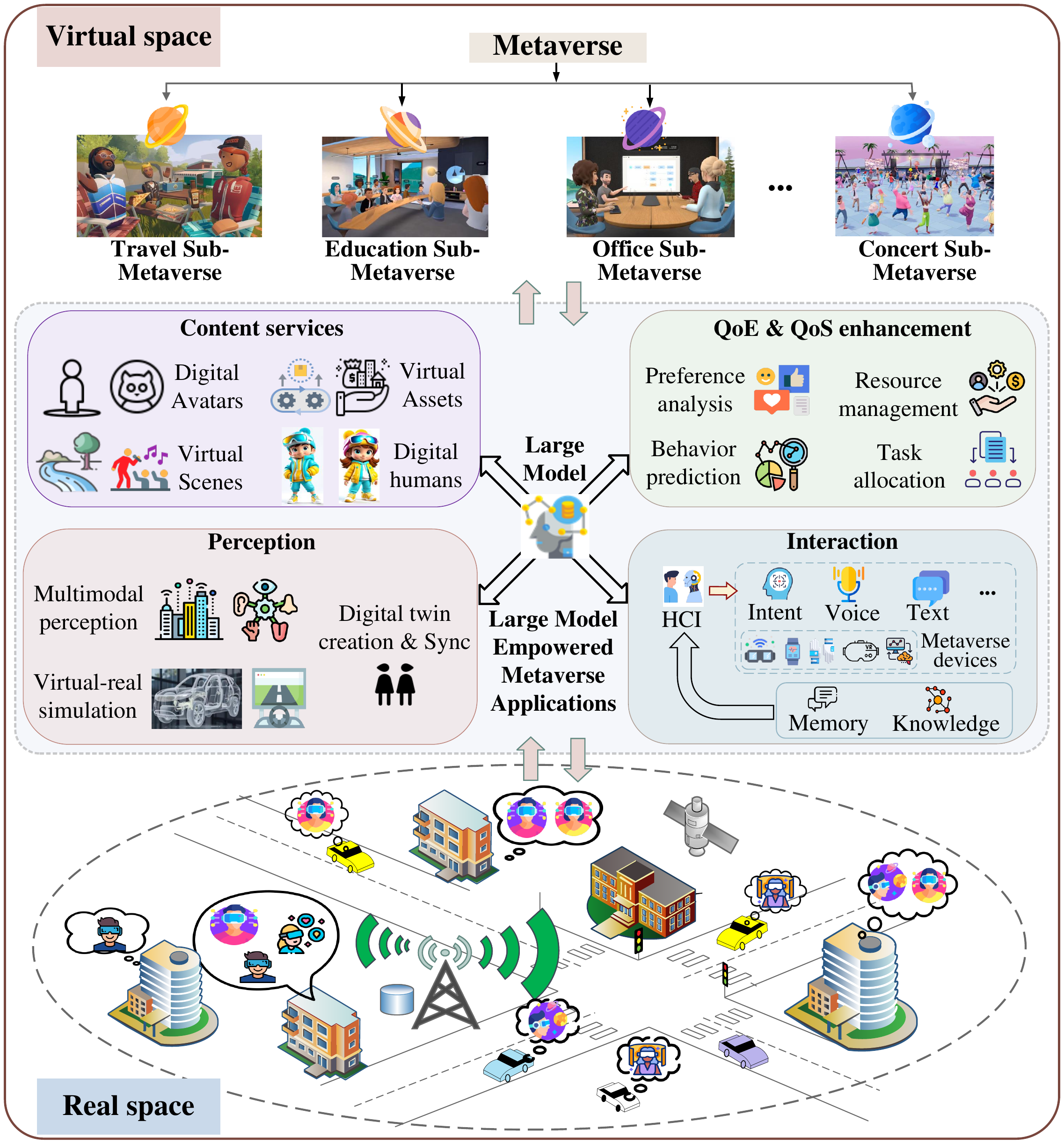}
  \caption{An overview of large model empowered Metaverse applications.}\label{fig:LM_Metaverse_Apps}\vspace{-3.5mm}
\end{figure}

\section{Overview and State-of-the-arts of Large Models Empowered Metaverse}\label{SOTA}
\subsection{Overview of Large Model Empowered Metaverse}\label{architecture}

Large models can be categorized into three main types \cite{Wang2023ChatGPT}: large language models (LLMs), large vision models (LVMs), and large multimodal models (LMMs).
\begin{itemize}
    \item \textit{LLMs} such as GPT4 focus on natural language processing (NLP) tasks such as text generation, translation, summarization, and question answering. 
    \item \textit{LVMs} such as DALL-E 3 specialize in visual tasks, such as image comprehension and text-to-image generation. 
    \item \textit{LMMs} such as Sora integrate multiple data modalities, such as text, images, and audio, to enhance cross-modal understanding and interaction. 
\end{itemize}

As depicted in Fig.~1, the key Metaverse applications empowered by large models include the following. 

\textit{1) Interaction:} Integrating LLMs and LMMs significantly enhances avatar-environment, avatar-avatar, and user-avatar interactions in the Metaverse. 
\begin{itemize}
    \item Natural language understanding: LLMs facilitate more natural and context-aware conversations within the Metaverse \cite{Wang2023ChatGPT}, enabling virtual environments to understand and respond to users more accurately.
    \item Emotional recognition: Through sentiment analysis and affective computing, large models can interpret emotional cues from voice and facial expressions \cite{Wang2024Human}, fostering empathetic and emotionally adaptive interactions for improved user satisfaction.
    \item Multimodal interfaces: LMMs allow for diverse input forms, including text, speech, gestures, and even brainwave signals, creating more immersive and engaging experiences.
    \item Intent understanding: By combining short-term memory with long-term user knowledge, LLMs can accurately interpret user behavior and intent, supporting adaptive, personalized interactions \cite{Lv2023GAImetaSurvey}. This enables avatars to mirror users' intentions and expressions, making social interactions more lifelike.
    \item Context awareness: LLMs enhance realism by adapting interactions to real-time environmental data, user history, and preferences \cite{Saddik2024Integration}, thus maintaining coherence and relevance in dialogue. This continuity is especially valuable for social and collaborative Metaverse applications.
\end{itemize}

\textit{2) Perception:} Large models enhance the Metaverse's capacity to understand and interpret real-world and simulated environments.
\begin{itemize}
    \item World recognition: LVMs and LMMs improve processing and understanding of visual and multimodal data, supporting the interpretation of both physical environments and simulated experiences within the Metaverse \cite{Xu2023Generative}.
    \item Digital twin modeling: These models enable highly accurate digital representations of real-world entities, environments, and behaviors \cite{Wang2023C3Meta}, enhancing the realism and fidelity of digital twins and virtual content in the Metaverse.
    \item Environmental adaptation: By analyzing contextual and sensory data, large models dynamically adjust virtual elements, aligning them with real-time changes in user surroundings for seamless mixed-reality experiences. 
\end{itemize}

\textit{3) Content creation:} Large generative models enable rapid, cost-effective production of high-quality, multimodal content, enriching the Metaverse with immersive digital elements. 
\begin{itemize}
    \item Avatar generation: Large models allow for highly customizable, lifelike avatars tailored to user preferences and applications \cite{Chamola2024Beyond}, enhancing personalization in virtual spaces.
    \item Virtual asset creation: Large models support the efficient generation of virtual goods and assets, including clothing, tools, and artwork \cite{Lv2023GAImetaSurvey}, expanding economic opportunities within the Metaverse.
    \item Virtual environment design: They enable the creation of complex digital landscapes, from realistic cityscapes to imaginative fantasy worlds, enriching the diversity and scope of Metaverse environments \cite{Wang2024Human}.
    \item Digital human creation: Large models also facilitate the creation of realistic digital humans who can serve as virtual customer service guides or personal assistants, providing more interactive and engaging user support.
\end{itemize}

\textit{4) QoE \& QoS enhancement:} Large models enhance QoE and QoS by enabling intelligent, user-centered system optimizations \cite{Wang2024Human} in the Metaverse.
\begin{itemize}
    \item User/avatar behavior prediction: Leveraging spatiotemporal correlations in historical behaviors, large generative models can anticipate user/avatar behavior patterns, enabling smoother virtual experiences \cite{Xu2023Generative}. 
    \item Task scheduling: Large discriminant models can dynamically allocate resources and schedule tasks based on demand, optimizing service quality and minimizing latency \cite{Huynh2023AIMetaSurvey}.
\end{itemize}

\begin{table*}[!t]
    \centering
    \caption{Academic and Industrial Advances of GAI and Large Model Empowered Metaverse}
    \begin{tabular}{|l|p{3.5cm}|p{3.7cm}|p{2.8cm}|p{2.55cm}|}
        \hline
        \textbf{Ref./Company} & \textbf{Benefit} & \textbf{Metaverse Application} & \textbf{Used GAI Model} & \textbf{Type of Large Model} \\
        \hline
        Huynh-The \textit{et al.} \cite{Huynh2023AIMetaSurvey} & Comprehensive AI potential in CV, NLP, digital twins, etc & Virtual world construction, gaming, healthcare, smart cities & Transformers, Diffusion, GANs, Autoencoders & {\ding{55}} \\
        \hline
        Chamola \textit{et al.} \cite{Chamola2024Beyond} & Enhances text, image, video, audio, 3D content generation & Content creation for Metaverse environments & GAI models & {\ding{55}} \\
        \hline
        Saddik \textit{et al.} \cite{Saddik2024Integration} & Medical content generation & Medical Metaverse & GPT-4 & LLM \\
        \hline
        Xu \textit{et al.} \cite{Xu2023Generative} & Realistic traffic simulations & Vehicular mixed-reality Metaverse & TSDreamBooth & {\ding{55}} \\
        \hline
        Raad \textit{et al.} \cite{Deepmind2024Scaling} & Language-driven agent adaptability & 3D environment interaction in games & SIMA & LLM \\
        \hline
        Microsoft & Enhanced avatar interaction & Mesh platform for virtual communication & Mesh-integrated AI & Multimodal \\
        \hline
        Google (Project Starline) & Realistic 3D communication & 3D virtual environments & GAI models& {\ding{55}} \\
        \hline
        OpenAI (DALL-E) & Digital asset creation & Artwork, environment designs in virtual worlds & DALL-E & LVM \\
        \hline
        Tencent & AI-driven digital humans & Virtual companions, digital human interaction & Digital human models & Multimodal \\
        \hline
        Meta (Horizon Worlds) & Personalized user experiences & Social VR interactions & Large language and vision models & Multimodal \\
        \hline
        NVIDIA (Omniverse) & Digital twin creation & Simulation, immersive experiences & GAI models & {\ding{55}} \\
        \hline
        Alibaba & Virtual shopping assistance & Interactive 3D shopping spaces & Avatar and assistant AI & Multimodal \\
        \hline
    \end{tabular}
    \label{tab:large_model_metaverse}
\end{table*}

\subsection{State-of-the-Arts}\label{SOTAs}

Large model technologies have emerged as a research hotspot for enhancing the Metaverse. 
Huynh-The \textit{et al.} \cite{Huynh2023AIMetaSurvey} comprehensively survey the potentials of AI in constructing virtual worlds in the Metaverse, covering computer vision (CV), NLP, networking, digital twin, blockchain, and neural interface. Besides, AI-driven Metaverse applications including gaming, manufacturing, healthcare, and smart cities are discussed. 
Particularly, GAI models including Transformers, stable diffusion, generative adversarial networks (GANs), autoencoders, and autoregressive models are highlighted for their transformative potential.
Chamola \textit{et al.} \cite{Chamola2024Beyond} discuss the use cases, technologies, and applications of GAI models in the Metaverse for tasks such as text, image, video, audio, and 3D object generation.
Saddik \textit{et al.} \cite{Saddik2024Integration} explore GPT-4 as a creative content generator in medical Metaverse environments, and examine its potential and limitations in integrating GPT-4 and medical Metaverse. 
Xu \textit{et al.} \cite{Xu2023Generative} leverage GAI for smart traffic simulations within a vehicular mixed-reality Metaverse by synthesizing data based on current location, historical trajectory, and user preferences. They also propose TSDreamBooth, a diffusion model generator for virtual traffic signs under various road conditions using the BelgiumTS dataset.
Raad \textit{et al.} \cite{Deepmind2024Scaling} introduce the scalable, instructable, multiworld agent (SIMA), which enables agents to follow natural language instructions across complex 3D environments, including research and commercial games. Their work demonstrates the versatility of language-guided AI across virtual settings, as LLM agents perform keyboard-and-mouse actions based on image and language inputs.

Leading technology companies are harnessing large model technologies to reshape Metaverse applications, fueling innovation and enhancing user engagement. Microsoft integrates AI into its Mesh platform\footnotemark[3], enabling AI-driven avatars for more interactive virtual experiences. Google's Project Starline\footnotemark[4] leverages GAI to construct realistic 3D environments and immersive communication spaces. OpenAI's DALL-E\footnotemark[5] powers the creation of unique digital assets, such as artwork and environmental designs, supporting creative customization in virtual worlds. Tencent focuses on AI-driven digital humans and virtual companions\footnotemark[6], enriching user interactions in virtual environments. Meta employs large models in Horizon Worlds\footnotemark[7] to generate new virtual worlds and personalize experiences, enhancing social VR interactions. NVIDIA's Omniverse platform\footnotemark[8] utilizes GAI to create digital twins, facilitating simulations and immersive experiences. Alibaba integrates AI-generated avatars and virtual assistants\footnotemark[9] to transform virtual shopping, guiding users in interactive 3D spaces.
\footnotetext[3]{\url{https://www.uctoday.com/collaboration/microsoft-to-introduce-ai-optimised-avatars-to-mesh-for-teams/}}
\footnotetext[4]{\url{https://starline.google/}}
\footnotetext[5]{\url{https://openai.com/index/dall-e-3/}}
\footnotetext[6]{\url{https://www.tencentcloud.com/products/ivh}}
\footnotetext[7]{\url{https://developers.meta.com/horizon}}
\footnotetext[8]{\url{https://www.nvidia.com/en-us/omniverse/}}
\footnotetext[9]{\url{https://www.alibabacloud.com/blog/alibaba-clouds-new-ai-tools-spotted-at-apsara-2023_600550}}

\section{Key Challenges of Large Models Empowered Metaverse}\label{challenges}

\subsection{Low Scalability in Virtual World Construction}\label{Challenge1}

Large models are essential for rendering large-scale, immersive, and highly realistic Metaverse scenes, but their enormous computational demands often result in significant scalability challenges. Current Metaverse platforms, such as Roblox, Fortnite, and Horizon Worlds, while attracting tens of millions of active users daily, struggle to support efficient interactions among large numbers of users. These platforms manage their vast user bases by dividing them into numerous smaller spaces or servers, where each server hosts only a limited number of users. For instance, Horizon Worlds can accommodate a maximum of 20 users interacting simultaneously per space and Fortnite's sandbox creative mode can only support 16 players per space\footnotemark[10], that falls far short of the vision of billions of participants in the future Metaverse.\footnotetext[10]{\url{https://www.designboom.com/technology/mark-zuckerberg-metaverse-horizon-worlds-avatar-virtual-reality-08-26-2022/}}

Essentially, rendering every detail of the virtual environment for each individual in the Metaverse is often unnecessary. By focusing on rendering user-specific interactions and prioritizing the rendering of relevant and engaging elements, the Metaverse can significantly enhance both scalability and user experience. Intention-aware and adaptive rendering, which tailors virtual content based on user preferences and intentions, not only reduces computational overhead but also ensures a more responsive and immersive experience, thereby enhancing both scalability and QoE in Metaverse.

\subsection{Constrained Responsiveness in Metaverse Rendering}\label{Challenge2} 
Current Metaverse rendering techniques often exhaustively render all potentially relevant scenes for each user without considering users' mobility patterns or specific intentions. This indiscriminate rendering leads to significant computational overhead and resource inefficiency, as many scenes are rendered but remain unused or unnoticed, ultimately degrading overall system performance. 

To improve efficiency, integrating user mobility analysis and intention prediction into rendering processes is essential. By employing advanced algorithms that track and analyze user movement patterns, the Metaverse can predict future actions and dynamically adjust rendering priorities to focus on the most relevant areas. For example, if an avatar's likely path and interactions are anticipated, the system can pre-render only those specific regions, thereby optimizing rendering resources and reducing latency. Additionally, such predictive rendering should be adaptive in real-time to dynamic user behaviors. 

\subsection{Low Adaptability in Dynamic Metaverse Environments}\label{Challenge3}
Metaverse platforms often face the challenge of rendering Metaverse environments using a one-size-fits-all approach, delivering the same scene quality and granularity to all users regardless of individual preferences, end device limitations, or network capabilities. This lack of adaptability reduces the overall QoE for users, as it fails to account for the dynamic nature of Metaverse interactions, such as users frequently entering and exiting different scenes and having varying levels of interest in specific virtual environments. 

It is essential to develop adaptive scene rendering techniques that can generate personalized and fine-grained content tailored to avatars' interests, preferences, and context.
For instance, avatars who prefer highly interactive elements can be provided with high-resolution, detailed scenes that enhance engagement, while those focused more on social interactions may receive simplified renderings that prioritize avatar clarity and response speed.

\section{Case Study: GAI-Based Metaverse Rendering Optimization}\label{solutions}

This section presents a case study on optimizing large model-empowered Metaverse rendering services, including a cloud-edge-end collaborative Metaverse architecture (in Sect.~\ref{method1}), mobility-aware Metaverse pre-rendering mechanism (in Sect.~\ref{method2}), and diffusion model based adaptive Metaverse rendering
strategy (in Sect.~\ref{method3}).

\begin{figure*}[!t]
\centering \setlength{\abovecaptionskip}{-0.1cm}
  \includegraphics[width=13.6cm]{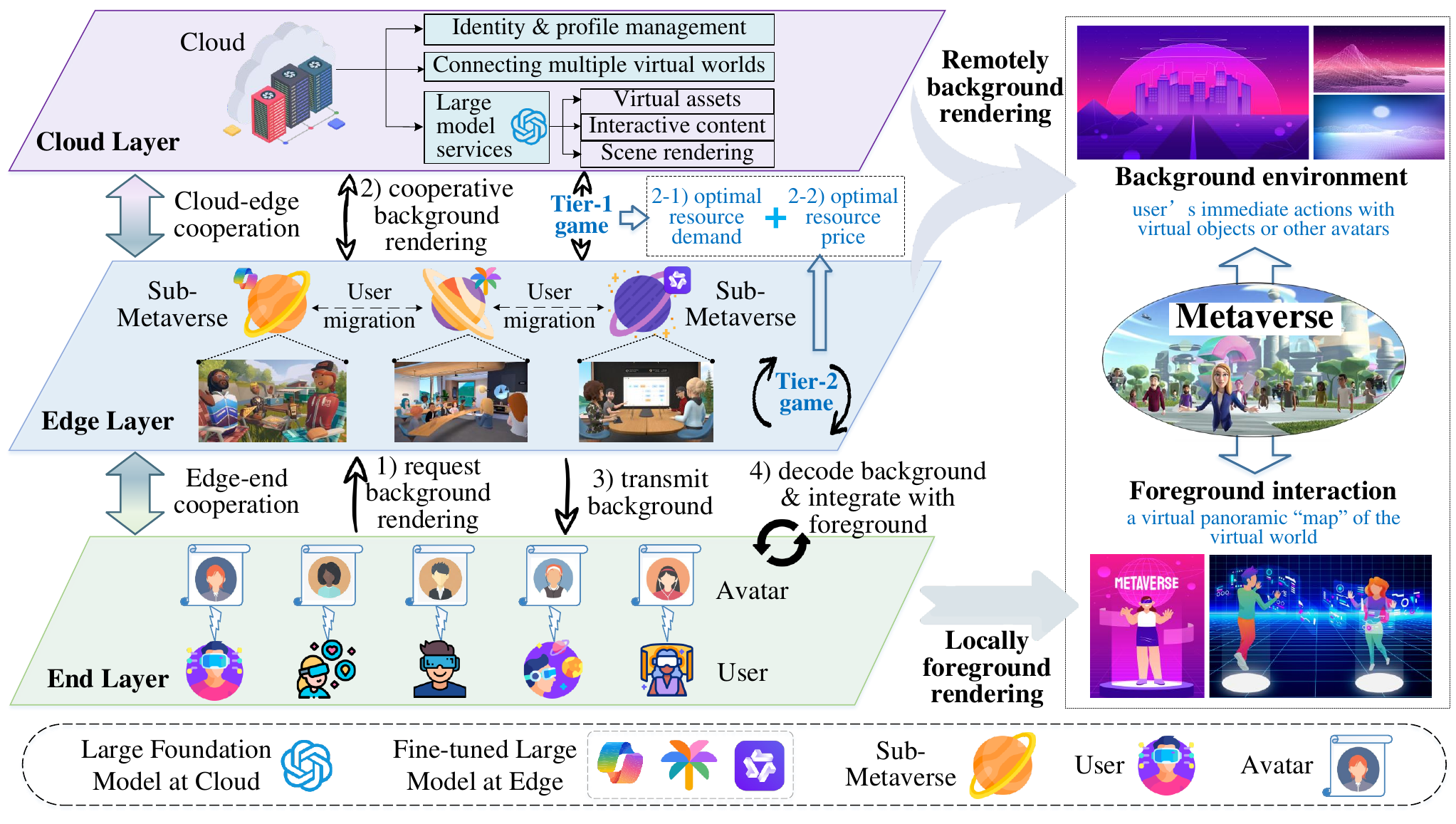}
  \caption{The cloud-edge-end three-layer architecture of Metaverse, which includes the cloud layer, the edge layer, and the end device layer.}\label{fig:Cloud_Edge_End_Metaverse}\vspace{-2.5mm}
\end{figure*}

\subsection{Cloud-Edge-End Collaborative Metaverse Architecture}\label{method1}

As shown in Fig.~\ref{fig:Cloud_Edge_End_Metaverse}, the cloud-edge-end collaborative Metaverse architecture can efficiently distribute rendering tasks across three layers, i.e., cloud servers, edge servers, and user devices, offering high-quality immersive experiences while minimizing latency and ensuring scalability. 

\begin{itemize}
    \item At the top layer, the cloud server serves as the primary storage hub for users' virtual assets, identities, and other critical data. The cloud's immense computational power supports large-scale data processing and storage, handling non-time-sensitive tasks and ensuring consistent global access to users' digital resources. 
    \item The middle layer consists of edge servers, which are geographically distributed and each manages a sub-Metaverse. Edge servers handle computational tasks that require low latency but need more processing power than the user's device can provide \cite{Wang2023C3Meta}. By rendering closer to the user, edge servers significantly reduce latency and improve the overall responsiveness of virtual experience. 
    \item At the bottom layer, users enter the Metaverse through devices such as VR headsets and head-mounted displays and known as ``avatars''. Although these devices have limited computational capabilities compared to cloud and edge servers, they are responsible for rendering real-time, low-latency interactions critical for immersion.
\end{itemize}

In the Metaverse, large model services are cooperatively delivered by the cloud and distributed edge nodes. While large foundational models on the cloud offer broad applicability, fine-tuning them through transfer learning, parameter-efficient fine-tuning, or quantification techniques \cite{Wang2023ChatGPT} on targeted datasets enables sub-Metaverses to create dedicated models tailored for specific applications, such as virtual classrooms or virtual dating. Furthermore, integrating outputs from these specialized large models on edge nodes back into the large foundational model on the cloud allows for the generation of richer, more nuanced responses. 
A typical Metaverse scene is divided into two components \cite{Lai2020C3Furion}: foreground interaction and background environment. 
\begin{itemize}
    \item \textit{Foreground interactions} refer to user's immediate actions, such as movements, gestures, and interactions with virtual objects or other avatars. They are less computationally intensive but require minimal latency to ensure smooth, responsive user experience. Typically, foreground interactions are rendered locally on the user's device, such as a VR headset, which allows for real-time, low-latency feedback that enhances the sense of immersion.
    \item \textit{Background environment} encompasses the broader virtual space where users engage in social activities, exploration, or interactions with dynamic surroundings. They often resembling a virtual panoramic ``map'' of the virtual world and require a larger rendering area with rich visual details, requiring extensive computational resources. 
\end{itemize}

\textit{System Workflow:} As depicted in Fig.~\ref{fig:Cloud_Edge_End_Metaverse}, when a user enters a Metaverse environment, their device handles the rendering of real-time interactions (step 1), such as interacting with other avatars or virtual objects. Simultaneously, large models deployed on the cloud and edge servers cooperatively render background maps (step 2). 
After that, the high-fidelity background is transmitted to user's device (step 3), where it synchronizes with the locally rendered foreground interactions (step 4) to create cohesive, immersive experience, based on user's current field of view (FoV). This process involves several steps: (i) the user requests background rendering, (ii) the cloud and edge collaborate to render and encode the panorama, (iii) the transmission to the user device, and (iv) the device decodes and integrates the background with the foreground interaction. 

We utilize a hierarchical game to analyze the optimal strategies of edge nodes and the cloud. The cloud's utility function is defined as the difference between its revenue from leasing resources and resource costs, while the utility function of edge node is defined as the difference between the resource satisfaction of rendering task and payment. Specifically, \textit{tier-1} is a sequential game between the cloud and edge nodes, while \textit{tier-2} is a non-cooperative game among edge nodes.
Using backward induction, we first analyze the optimal rendering resource demand strategy for each edge node by calculating the best response functions of all other edge nodes, which leads to the Nash equilibrium strategy for each edge node in this non-cooperative game (step 2-1). Then, by substituting optimal strategies of edge nodes into the cloud's utility function, we derive the cloud's optimal resource pricing strategy via an iterative algorithm based on gradient descent due to the difficulty in obtaining its explicit expression (step 2-2).

\begin{figure*}[!t]
\centering \setlength{\abovecaptionskip}{-0.cm}
  \includegraphics[width=15.5cm]{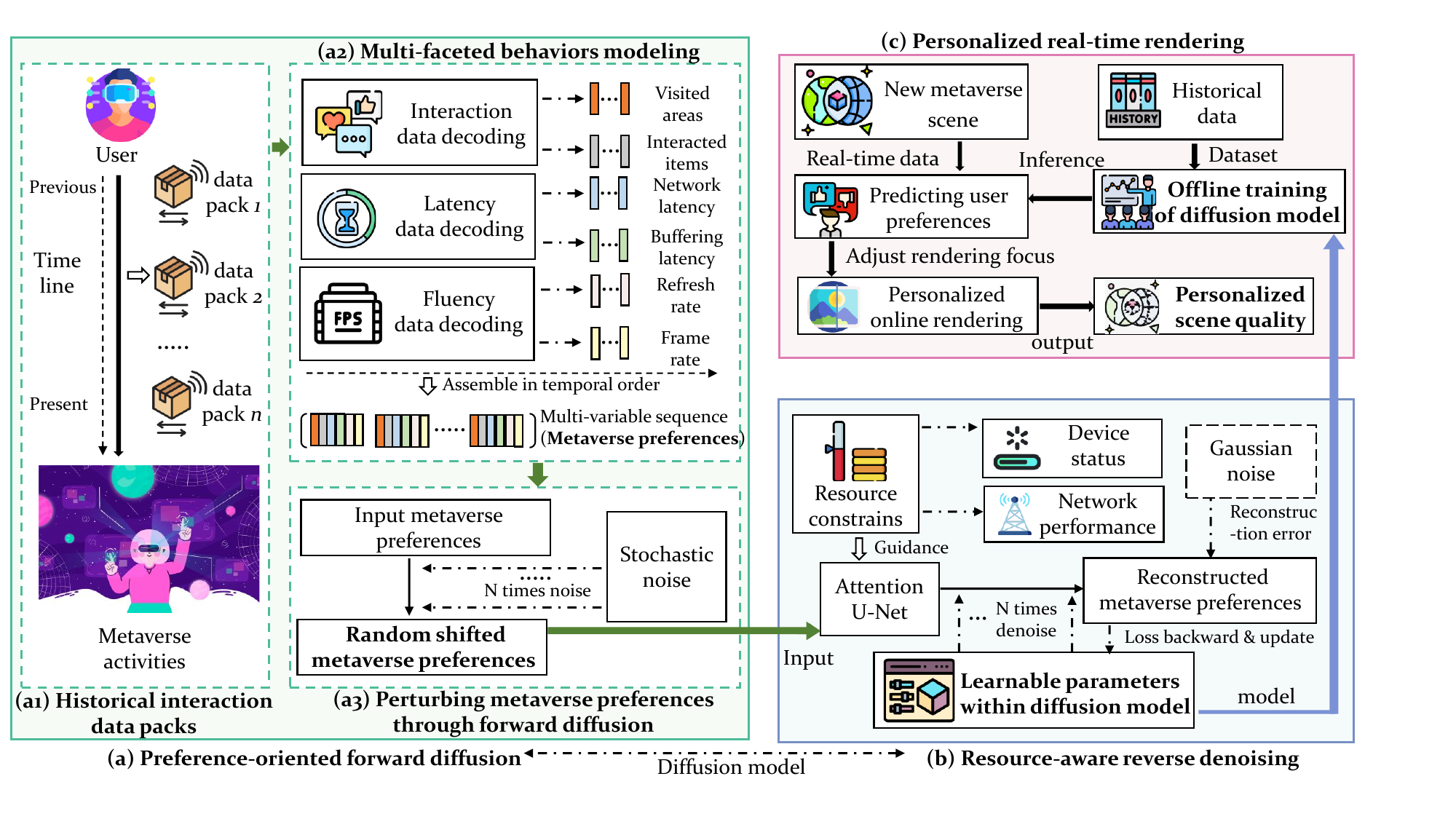}
  \caption{Illustration of diffusion model based personalized Metaverse rendering strategy including: (a) preference-oriented forward diffusion, (b) resource-aware reverse denoising, and (c) personalized real-time rendering.}\label{fig:Diffusion_Rendering}\vspace{-2.5mm}
\end{figure*}

\subsection{Mobility-Aware Metaverse Pre-Rendering Mechanism}\label{method2}
Rendering the background environment in the Metaverse involves the above four steps, which can be time-consuming and decrease frame rates. Additionally, misalignment between rendered scenes and users' movements can negatively impact overall QoE. 
To tackle these challenges, we employ a pre-rendering technique that utilizes panorama compression while leveraging user mobility and spatial similarity.

\textit{Phase 1: Mobility-aware pre-rendering.} By discretizing the virtual environment into dense grid points, the system pre-renders panoramas of neighboring points, based on anticipated avatar movements. Each grid point corresponds to a specific location, and when an avatar moves, the system retrieves and displays the pre-rendered scene of the nearest neighboring point. 
Given the unpredictability of avatar movement, the system continuously renders and encodes panoramas for adjacent grid points at each time step. By the time the avatar shifts position, the panoramas are ready for immediate decoding and integration with locally rendered foreground interactions. 
Note that the total time for background rendering requests, collaborative rendering \& encoding, and transmission to the user should not exceed the time \textit{{T}} for a user to move from one grid point to an adjacent one.
As such, it ensures a seamless, immersive experience by maintaining high visual quality and minimizing lag.

\textit{Phase 2: Spatial panorama compression.} Due to the dense packing of grid points, nearby panoramas often exhibit significant spatial similarity. This characteristic motivates us to encode and compress panorama data without sacrificing quality. Our scene segmentation and compression strategy divides the virtual environment into distinct regions, where the panorama at the central grid point of each region is encoded as an I-frame. The I-frame serves as a reference that can be decoded independently, while the panoramas of surrounding grid points are encoded as P-frames, relying on the I-frame for decoding. Users store all I-frames along with essential information about the P-frames, such as virtual object dimensions. Here, too high a density can cause rendering discontinuities during movement, while too low a density may impose excessive rendering burdens. Research indicates that an optimal density is approximately 2cm in the real world \cite{Boos2016Flashback}. For instance, if a user moves at a speed of 1 m/s, the unit time \textit{{T}} for processing transitions would be around 20 ms.

\subsection{Diffusion Model Based Personalized Metaverse Rendering Strategy}\label{method3}

Generally, for each Metaverse user, the rendering quality is constrained by individual preferences, end device capabilities, and real-time network conditions. To address adaptability issues, we devise a diffusion model based rendering strategy including the following three phases, as illustrated in Fig.~\ref{fig:Diffusion_Rendering}.

\textit{Phase 1: Preference-oriented forward diffusion.} As depicted in Fig.~\ref{fig:Diffusion_Rendering}(a), we first model user's multi-faceted behaviors in the Metaverse, with the historical behavior sequence representing his/her Metaverse preferences. Specifically, we decode data packets from the user's previous Metaverse activities, assembling interaction data (e.g., visited areas and interacted items), latency data (e.g., network and buffering latency), and fluency data (e.g., refresh rate and frame rate) into a multi-variable sequence 
\textit{M}, representing the Meteverse user's preferences. Next, we apply a forward diffusion process to incrementally perturb \textit{M} into \textit{M'} via controlled transformations, adding stochastic noise to simulate potential shifts in user interests and context. Each iteration introduces randomness, enabling the diffusion model to explore a broad range of potential preferences.

\textit{Phase 2: Resource-aware reverse denoising.} As shown in Fig.~\ref{fig:Diffusion_Rendering}(b), the reverse denosing phase is designed to progressively reconstruct  user's original preferences \textit{M} from the distorted preferences \textit{M'} generated during the forward diffusion phase, learning to forecast user's preferences when rendering new scenes in the Meteverse. As user's preferences in scene quality is often constrained by device status (e.g., CPU/GPU load, battery level) and network performance(e.g., bandwidth fluctuation), we set up resource constraints \textit{S} by monitoring user's device status and network performance in real-time. For each reverse denosing iteration, Attention U-Net \cite{AttentionUNet} is adopted to learn the conditional probabilities under resource constraints \textit{S}, where \textit{S} serves as the additional input to guide the reverse denoising of \textit{M'}, allowing the diffusion model to adaptively learn user's preferences based on resource constraints.

\textit{Phase 3: Personalized real-time rendering.} As illustrated in Fig.~\ref{fig:Diffusion_Rendering}(c), {our proposed diffusion scheme adopts a hybrid offline-online rendering strategy. It} first utilizes user's historical data to conduct offline training for preferences learning, then leverages Metaverse real-time data to perform online rendering tailed to user's preferences and resource constraints. During offline training, we aim to predict noise of user's distorted preferences \textit{M'} in the reverse denoising phase and minimizes the mean square error between Gaussian noise and the predicted ones {with step-by-step error analysis, ensuring effectiveness and preference alignment.} 

After offline training, the diffusion model can conduct online rendering. For a given new scene, we perform forward diffusion {followed by skip-step} reverse denoising to reconstruct the interaction probabilities of non-interacted items in the new scene, {thereby expediting the inference process of the trained diffusion model for real-time personalized rendering.}
By utilizing these predicted interaction probabilities, we dynamically adjust the rendering focus in real time, ensuring high-quality visuals in areas preferred by the user while adapting scene details to align with personalized interests and resource constraints.

\subsection{Performance Evaluation}\label{Evaluation}
We conduct simulations to validate the effectiveness of the proposed Metaverse rendering optimization system.
We implement the diffusion model with the Attention U-Net architecture \cite{AttentionUNet}, configured with 512 dimensions and 8 attention heads.  
For the training phase, we employ the Adam optimizer with an initial learning rate of 0.0001 and a batch size of 256, training for 20 epochs with early stopping to prevent overfitting. The diffusion steps are set to 700, with a diffusion variance schedule coefficient linearly increasing from 0.0001 to 0.04. For the validation phase, our prototype is implemented on a cloud server running Ubuntu 20.04 OS, equipped with a 2.60 GHz Intel Xeon CPU, 256 GB of memory, and dual NVIDIA RTX 3090 GPUs. We apply two mainstream optimization strategies, i.e., Markov Decision Processes (MDP) and Random Optimization (RO) as benchmarks for performance comparison. 
{All the validation benchmarks are using the optimal parameters according to their respective literature. For the MDP baseline, the rendering policy is optimized using value iteration with a discount factor of 0.95, and for the RO baseline, 21 rendering parameter combinations are randomly sampled per scene, with the best-performing configuration selected.}

\begin{figure}[ht]
\fbox{\centering\setlength{\abovecaptionskip}{-0.1cm}
\includegraphics[width=7cm]{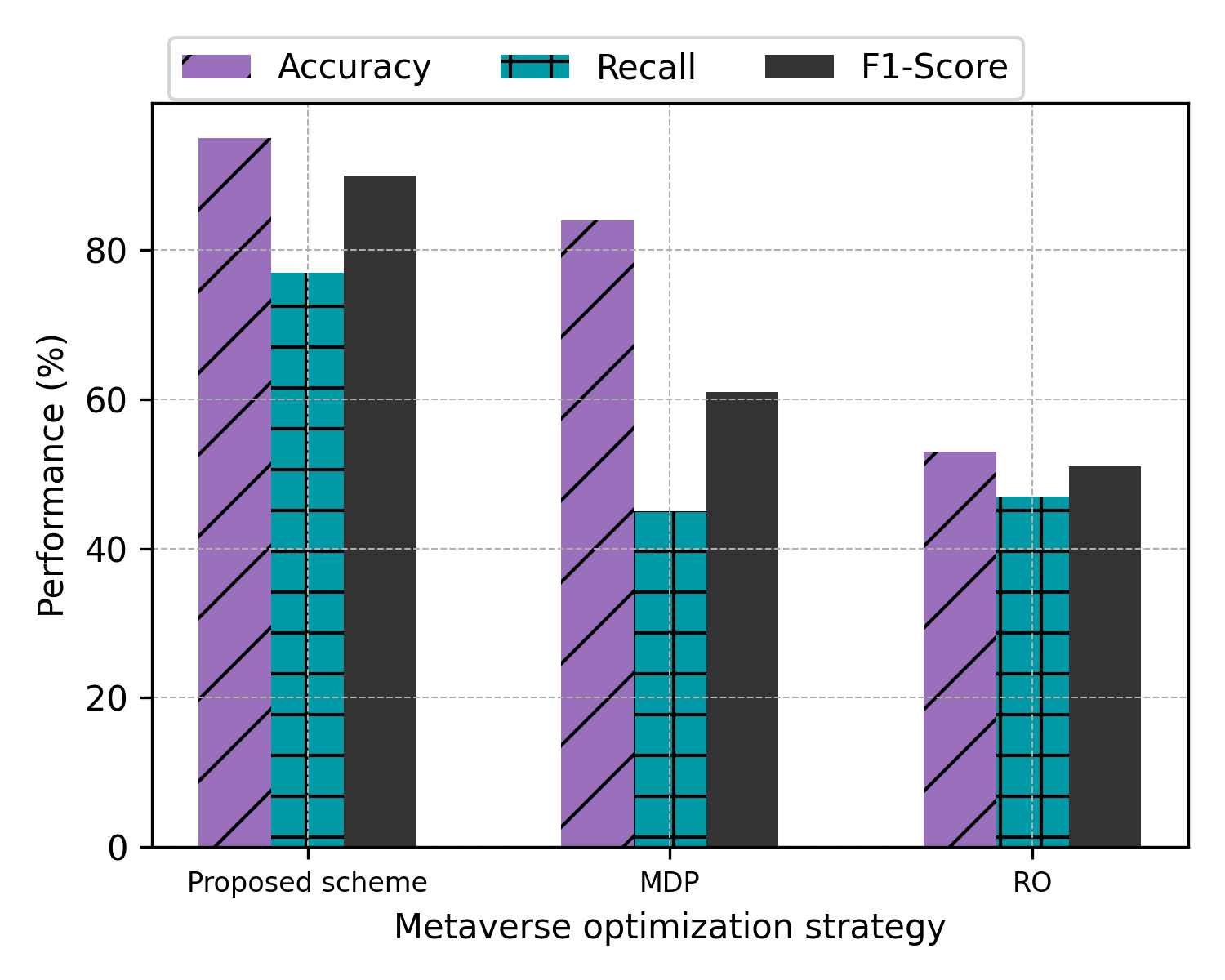}}
\caption{Comparison of Metaverse rendering optimization performance in the proposed scheme and benchmarks.}
\label{evaluation:performance}\vspace{-2.5mm}
\end{figure}

\begin{figure}[ht]
\fbox{\centering\setlength{\abovecaptionskip}{-0.1cm}
\includegraphics[width=7cm]{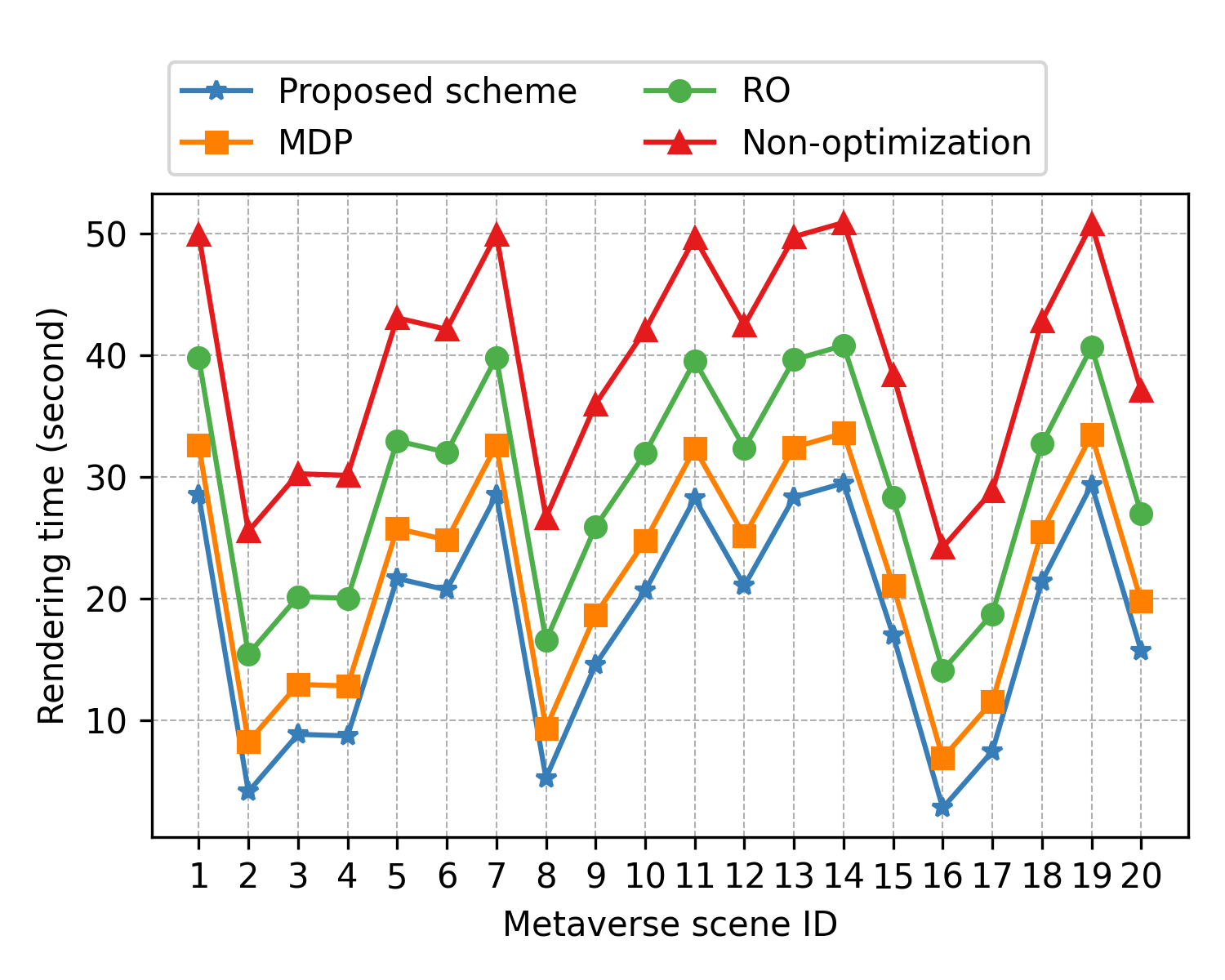}}
\caption{Comparison of computation overhead of Metaverse rendering in the proposed scheme and benchmarks.}
\label{evaluation:overhead}\vspace{-2.5mm}
\end{figure}

We first evaluate the Metaverse rendering optimization performance in different schemes. As depicted in Fig.~\ref{evaluation:performance}, the proposed scheme outperforms the benchmarks in accuracy, recall, and F1-Score. Specifically, our proposed scheme reaches the following improvements over the best-performing benchmark: a 13.1\% increase in optimization accuracy, a 71.11\% boost in optimization recall, and a 47.54\% raise in optimization F1-Score. These improvements in optimization performance demonstrate the effectiveness of our proposed scheme in predicting users' intentions and preferences in Metaverse, adapting to dynamic and personalized rendering requirements, thereby enhancing overall QoS and QoE for Metaverse users.

Next, we evaluate the rendering time of {20 one-minute Metaverse scenes, each comprising 3,600 frames at 60 FPS,} on dual NVIDIA RTX 3090 GPUs in different schemes. 
As observed in Fig.~\ref{evaluation:overhead}, under non-optimization scheme, the Metaverse rendering overhead incurs an average of 39.54 second in 20 different Metaverse scenes. By contrast, the proposed scheme reduces the rendering overhead to an average of 18.14 second, which achieves the best rendering efficiency among all benchmarks, reducing rendering time by approximately 18.44\% compared to MDP, 38.38\% compared to RO, and 54.12\% compared to non-optimization.

\section{Future Research Directions}\label{future}
This section discusses open problems that remain to be investigated in the field of large model empowered Metaverse.

\subsection{Collaborative Large Model Agents for Metaverse Optimization}
Large model agents, i.e., AI-driven entities powered by large models, can work together to manage complex tasks such as resource allocation, user behavior prediction, and content adaptation, thereby significantly improving the efficiency and interactivity of Metaverse environments \cite{Deepmind2024Scaling}. 
For instance, agents can jointly adjust virtual scenes based on real-time analytics, optimizing graphics rendering, latency, and user interactions to maintain a seamless experience. By leveraging technologies such as multi-agent reinforcement learning and federated learning, these agents can share insights without compromising data privacy, continuously learning and adapting to evolving user demands. 
However, coordinating multiple agents presents challenges, such as ensuring reliable communication and avoiding conflicts in decision-making, which could disrupt user experiences. Additionally, the high computational demands of these systems require efficient energy management strategies to prevent excessive power consumption. 

\subsection{Semantic Communications in Large Model Empowered Metaverse}
Semantic communications hold great potential in enhancing Metaverse by enabling more efficient and context-aware information exchange. In large model empowered Metaverse, semantic communication can interpret and transmit the most relevant information, such as user intentions or contextual cues, rather than raw data \cite{Wang2023SemanticMeta}. For example, by leveraging large models and knowledge graphs, semantic communication systems can prioritize and compress data based on its significance to the current virtual context, improving the responsiveness of virtual interactions and reducing network bandwidth requirements. This allows for smoother and more immersive experiences, even in complex and data-intensive scenarios. 
However, implementing semantic communications presents challenges, such as accurately interpreting relevant semantics and robust error-handling mechanisms in diverse and unpredictable Metaverse environments. 

\subsection{Security and Privacy in Large Model Empowered Metaverse}
As large models become integral to the Metaverse, addressing security and privacy concerns is crucial to maintaining a secure and trustworthy Metaverse \cite{Awadallah2024AImetaSecSurvey}. 
One major risk is hallucination, where large models generate false or misleading information that could disrupt user experiences or propagate misinformation. Another concern is the jailbreak, where malicious actors exploit vulnerabilities to manipulate model behavior, potentially bypassing safety protocols and introducing harmful content. Additionally, data memorization poses a significant privacy threat, as models inadvertently retain sensitive user data, increasing the risk of unauthorized data exposure. 
Key challenges in mitigating these risks include developing robust defense mechanisms to detect and prevent manipulations without degrading model performance. Another challenge is designing new privacy preservation method to ensure data privacy during model training and deployment processes without sacrificing efficiency.

\section{Conclusion}\label{conclusion}
Large models, including LLMs and LVMs, bring significant advancements to the Metaverse by enabling precise real-world modeling, rich content generation, lifelike avatars, and multimodal interactions, which together elevate QoE and QoS. 
This paper has reviewed recent academic and industrial advancements in integrating large models within the Metaverse, highlighting persistent challenges, such as scalability limitations, low adaptability, and reduced responsiveness in dynamic environments. 
To address these issues, we have presented a case study on Metaverse rendering optimization featuring a cloud-edge-end collaborative model, a mobility-aware pre-rendering mechanism, and an adaptive rendering strategy utilizing diffusion models. Extensive evaluations of our prototype have validated that our approach improves rendering efficiency and reduces system overheads. 
This work contributes to the development of sustainable, adaptive Metaverse systems and is expected to shed more light on ongoing efforts for advancing future immersive virtual worlds.

\section*{Acknowledgement}\label{sec:Acknowledge}
This work was supported in part by NSFC under Grant 62302387, Grant
U22A2029, Grant U24A20237, Grant 62302288, Grant 62402379, Grant 62371280.

\vspace{-1.cm}
\begin{IEEEbiographynophoto}{Yuntao Wang}
is currently an Assistant Professor with the School of Cyber Science and Engineering in Xi'an Jiaotong University, China. His research interests include security and privacy in UAV networks and LLM agents.
\end{IEEEbiographynophoto}

\begin{IEEEbiographynophoto}{Qinnan Hu}
is working on the Ph.D degree with the school of Cyber Science and Engineering of Xi'an Jiaotong University, China. His research interests include blockchain system security.
\end{IEEEbiographynophoto}

\begin{IEEEbiographynophoto}{Zhou Su}
is a professor with Xi'an Jiaotong University and his research interests include multimedia communication, wireless communication, network security and network traffic. Dr. Su has published technical papers, including top journals and top conferences, such as IEEE JSAC, IEEE/ACM ToN, IEEE TWC, IEEE INFOCOM, etc. He received the Best Paper Award of International Conference IEEE AIoT2024, IEEE WCNC2023, IEEE VTC-Fall2023, IEEE ICC2020, etc. 
He is an Associate Editor of {\scshape IEEE Internet of Things Journal}, and {\scshape IEEE Open Journal of the Computer Society}. He is the chair of IEEE VTS Xi'an Chapter Section.
\end{IEEEbiographynophoto}

\begin{IEEEbiographynophoto}{Linkang Du} 
received his B.E. and Ph.D. degrees from Zhejiang University in 2018 and 2023, respectively. He 
is currently an assistant professor at the School of Cyber Science and Engineering, Xi'an Jiaotong University, Xi'an, China. His research interests include data privacy protection and trustworthy machine learning.
\end{IEEEbiographynophoto}

\begin{IEEEbiographynophoto}{Qichao Xu} received Ph.D degree from the school of
Mechatronic Engineering and Automation, Shanghai
University, Shanghai, China, in 2019. He is
an Associate Professor with Shanghai university. His
research interests are in trust and security, the general
area of wireless network architecture, internet of
things, vehicular networks, and resource allocation.
\end{IEEEbiographynophoto}

\begin{IEEEbiographynophoto}{
Weiwei Li} received the Ph.D. degree from the school of Mechatronic Engineering and Automation of Shanghai University, Shanghai, China, in 2022. She is currently a lecturer with the College of Computer Science and Technology with Shanghai University of Electric Power, Shanghai, China. Her research interests are in the general area of cyber security and wireless network architecture.
\end{IEEEbiographynophoto}


\begin{thebibliography}{1}

\bibitem{Wang2023MetaverseSurvey}
M. Xu \textit{et al.}, ``A Full Dive Into Realizing the Edge-Enabled Metaverse: Visions, Enabling Technologies, and Challenges'' \emph{IEEE Communications Surveys \& Tutorials}, vol. 25, no. 1, pp. 656--700, 2023.


\bibitem{Chamola2024Beyond}
V. Chamola \textit{et al.}, ``Beyond Reality: The Pivotal Role of Generative AI in the Metaverse,'' \emph{IEEE Internet of Things Magazine}, vol. 7, no. 4, pp. 126--135, July 2024.


\bibitem{Wang2023ChatGPT}
Y. Wang, Y. Pan, M. Yan, Z. Su and T. H. Luan, ``A Survey on ChatGPT: AI–Generated Contents, Challenges, and Solutions,'' \emph{IEEE Open Journal of the Computer Society}, vol. 4, pp. 280-302, 2023.


\bibitem{Xu2023Generative}
M. Xu \textit{et al.}, ``Generative AI-Empowered Simulation for Autonomous Driving in Vehicular Mixed Reality Metaverses,'' \emph{IEEE Journal of Selected Topics in Signal Processing}, vol. 17, no. 5, pp. 1064--1079, Sept. 2023.

\bibitem{Saddik2024Integration}
A. E. Saddik and S. Ghaboura, ``The Integration of ChatGPT With the Metaverse for Medical Consultations,'' \emph{IEEE Consumer Electronics Magazine}, vol. 13, no. 3, pp. 6--15, May 2024.


\bibitem{Deepmind2024Scaling}
M. A. Raad \textit{et al.}, ``Scaling Instructable Agents Across Many Simulated Worlds,'' \emph{arXiv preprint arXiv:2404.10179}, pp. 1--31, 2024.

\bibitem{Huynh2023AIMetaSurvey}
T. Huynh-The, Q.-V. Pham, X.-Q. Pham, T. T. Nguyen, Z. Han, D.-S. Kim, ``Artificial Intelligence for the Metaverse: A Survey,'' \emph{Engineering Applications of Artificial Intelligence}, vol. 117, p. 105581, 2023.

\bibitem{Lv2023GAImetaSurvey}
Z. Lv, ``Generative Artificial Intelligence in the Metaverse Era,'' \emph{Cognitive Robotics}, vol. 3, pp. 208-217, 2023.

\bibitem{Wang2024Human}
Y. Wang, L. Wang, K. L. Siau, ``Human-Centered Interaction in Virtual Worlds: A New Era of Generative Artificial Intelligence and Metaverse,'' \emph{International Journal of Human–Computer Interaction}, pp. 1-43, 2024.


\bibitem{Wang2023C3Meta}
R. Wang, J. Wang, Y. Hao, L. Hu, S. A. Alqahtani and M. Chen, ``C3Meta: A Context-Aware Cloud-Edge-End Collaboration Framework toward Green Metaverse,'' \emph{IEEE Wireless Communications}, vol. 30, no. 5, pp. 144-150, 2023.

\bibitem{Lai2020C3Furion}
Z. Lai \textit{et al.}, ``Furion: Engineering High-Quality Immersive Virtual Reality on Today's Mobile Devices,'' \emph{IEEE Transactions on Mobile Computing}, vol. 19, no. 7, pp. 1586-1602, 2020.

\bibitem{Boos2016Flashback}
K. Boos, D. Chu and E. Cuervo, ``Flashback: Immersive Virtual Reality on Mobile Devices via Rendering Memoization,'' \emph{Annual International Conference on Mobile Systems, Applications, and Services (MobiSys'16)}, 2016, pp. 291–304.

\bibitem{AttentionUNet}
O. Oktay \textit{et al.}, ``Attention U-Net:
Learning Where to Look for the Pancreas,'' \emph{arXiv preprint arXiv:1804.03999}, pp. 1-11, 2018.

\bibitem{Wang2023SemanticMeta}
Y. Huang \textit{et al.}, ``ISCom: Interest-Aware Semantic Communication Scheme for Point Cloud Video Streaming on Metaverse XR Devices,'' \emph{IEEE Journal on Selected Areas in Communications}, vol. 42, no. 4, pp. 1003-1021, Apr. 2024.


\bibitem{Awadallah2024AImetaSecSurvey}
A. Awadallah \textit{et al.}, ``Artificial Intelligence-Based Cybersecurity for the Metaverse: Research Challenges and Opportunities,'' \emph{IEEE Communications Surveys \& Tutorials}, 2024, doi: 10.1109/COMST.2024.3442475.


\end{thebibliography}
\end{document}